# Rydberg series of dark excitons and the conduction band spin-orbit splitting in monolayer WSe$_2$

P. Kapuściński[1,2,*], A. Delhomme[1], D. Vaclavkova[1], A. O. Slobodeniuk[3], M. Grzeszczyk[4], M. Bartos[5], K. Watanabe[6], T. Taniguchi[7], C. Faugeras[1] and M. Potemski[1,4,*]

[1]*Laboratoire National des Champs Magnétiques Intenses, CNRS-Univ. Grenoble Alpes-UPS-INSA-EMFL, 25, avenue des Martyrs, 38042 Grenoble, France*
[2]*Department of Experimental Physics, Wrocław University of Technology, Wybrzeże Wyspiańskiego 27, 50-370 Wrocław, Poland*
[3]*Department of Condensed Matter Physics, Faculty of Mathematics and Physics, Charles University, Ke Karlovu 5, Praha 2 CZ-121 16, Czech Republic*
[4]*Institute of Experimental Physics, Faculty of Physics, University of Warsaw, ul. Pasteura 5, 02-093 Warszawa, Poland*
[5]*Central European Institute of Technology, Brno University of Technology, Purkyňova 656/123, 612 00 Brno, Czech Republic*
[6]*Research Center for Functional Materials, National Institute for Materials Science, 1-1 Namiki, Tsukuba 305-0044, Japan*
[7]*International Center for Materials Nanoarchitectonics, National Institute for Materials Science, 1-1 Namiki, Tsukuba 305-0044, Japan*
*Corresponding authors: piotr.kapuscinski@lncmi.cnrs.fr, marek.potemski@lncmi.cnrs.fr

Abstract

Strong Coulomb correlations together with multi-valley electronic bands in the presence of spin-orbit interaction and possible new optoelectronic applications are at the heart of studies of the rich physics of excitons in semiconductor structures made of monolayers of transition metal dichalcogenides (TMD). In intrinsic TMD monolayers the basic, intravalley excitons are formed by a hole from the top of the valence band and an electron from either the lower or upper spin-orbit-split conduction band subbands: one of these excitons is optically active, the second one is "dark", although possibly observed under special conditions. Here we demonstrate the s-series of Rydberg dark exciton states in monolayer WSe$_2$, which appears in addition to a conventional bright exciton series in photoluminescence spectra measured in high in-plane magnetic fields. The comparison of energy ladders of bright and dark Rydberg excitons is shown to be a method to experimentally evaluate one of the missing band parameters in TMD monolayers: the amplitude of the spin-orbit splitting of the conduction band.

Introduction

In most of low-dimensional, direct-gap semiconductor systems (quantum wells and monolayers of transition metal dichalcogenides, quantum wires and dots), the near band edge electron-hole excitations, the excitonic states, can be classified into two categories depending whether their associated projection of the total angular momentum is $j_z=\pm 1$ or it is $j_z=\pm 2$ or 0. According to standard selection rules for optical transitions, the excitons with $j_z=\pm 1$ can be excited by light and/or can recombine by emitting photons and are therefore referred to as bright excitons. Instead, excitons with $j_z=\pm 2$ or 0 are termed dark as they do not efficiently couple to the radiation field. Although dark, the $j_z=\pm 2$ or $j_z=0$ excitons can be brightened under special conditions, e.g., under application of the magnetic field in an appropriate configuration[1,2] and thus possibly initiating the photoluminescence signal (though still hardly observed in absorption-type spectra). Dark excitons can largely alter the dynamics of optical excitations and, as long-lived bosonic quasiparticles, are possible candidates to form a condensate exciton phase[3]. Those states have been the object of vast investigations in semiconductor quantum wells, dots and wires, in carbon nanotubes and more recently in monolayer semiconductors[4-19]. All these studies have been however largely focused on excitonic ground states whereas the physics of dark excitons associated with the excited excitonic states is a little explored area[20–22]. This, in particular, applies to two-dimensional semiconductors which often display a characteristic series of Rydberg $ns$ states of bright excitons[23–26] whereas a demonstration of the corresponding series of dark excitonic states is missing so far.

In this paper, we report on low-temperature magneto-optical study of Rydberg series of excitonic states as they appear in a WSe$_2$ monolayer. Notably, in this representative example of a 2D semiconductor, the rich spectrum of excited excitonic states can be relatively easily traced with luminescence experiments[25,27–29]. Profiting of this fact we brighten dark excitons in our sample by applying an in-plane magnetic field and uncover their $ns$ ($n$=1, up to 4) Rydberg series which appears in addition to a Rydberg spectrum of bright excitons[24,25]. By comparing energy diagrams of dark and bright exciton series, we experimentally derive the single particle separation between spin-orbit-split conduction band subbands in a WSe$_2$ monolayer. Intriguingly, this separation is found to be of about two times smaller than that commonly assumed from theoretical modelling of electronic bands of monolayer semiconductors.

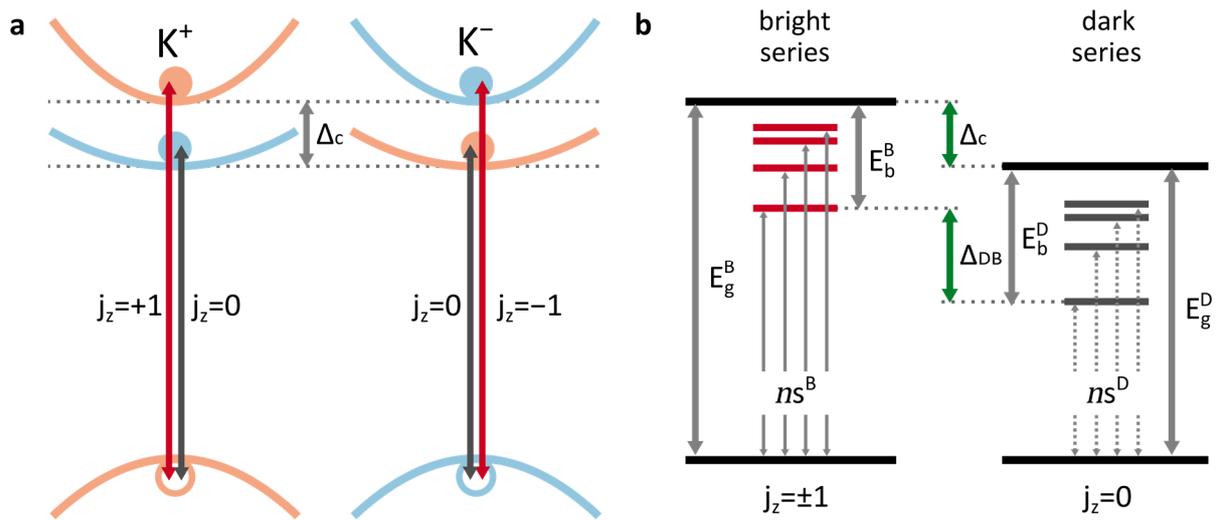

*Figure 1. **a** Schematic representation of the interband transitions in a single particle picture. The spin-up and spin-down states are shown as light orange and light blue parabolas, respectively. Optically active (bright) transitions at K$^+$ and K$^-$ valleys are represented by the red arrows, while optically inactive (dark) transitions are depicted in dark grey. The light grey arrow represents the conduction band spin-orbit splitting ($\Delta_c$). **b** The energy diagram of excitonic transitions: Rydberg series of bright ($ns^B$) and dark ($ns^D$) resonances. The horizontal red and dark grey lines represent bright and dark excitonic levels, respectively. The light grey arrows represent the band gaps for bright and dark excitons ($E_g^{B/D}$) and the binding energies of the ground states of bright and dark excitons ($E_b^{B/D}$). The energy splitting between ground states of bright and dark excitons ($\Delta_{DB}$) and the conduction band spin-orbit splitting ($\Delta_c$) are represented by the green arrows.*

The focus of our study can be clearly read from Fig.1. Fig.1a illustrates a simple generic scheme of the band structure of a WSe$_2$ monolayer, as it is found in a close vicinity of the direct bandgap at the two K$^+$ and K$^-$ points of the Brillouin zone. We consider only three relevant bands (per each K$^+$ and K$^-$ point): two spin-orbit split subbands in the conduction band (weakly separated in energy, by $\Delta_c \sim$ tens of meVs) but only one, top valence subband (neglecting the lower one separated by hundreds of meVs from the upper subband). As generally accepted and shown in Fig.1a, the interband, intravalley transitions which involve the electronic states of the upper conduction subband are optically allowed whereas those associated with lower subband are characterized by j$_z$=0 and thus are not optically active under conventional conditions/configurations[15,30]. The j$_z$=0 transitions are weakly allowed only in the configuration for which the electromagnetic wave has its k-vector aligned along the monolayer plane and non-zero out of plane component of the electric field, or in the presence of an in-plane magnetic field. Please, note, that in the case of TMD monolayers, the j$_z$=±1 selection rules for optical transitions coincide with the ansatz of conservation of electronic spins in optical transitions. The band to band image of optical transitions is not, however, sufficient to describe the optical response of semiconductors, and in particular that of the investigated structure: the electron-hole Coulomb interaction and hence the excitonic effects have to be taken into account. As shown in Fig. 1b, one may expect the appearance of two distinct $ns$ Rydberg series of near band edge excitons in our WSe$_2$ monolayer: the routinely observed bright exciton series[23–26] associated with the upper conduction band subband, but also the dark exciton series associated with the lower conduction band subband, revealed in the present report. In the limit of large $n$, the energy difference between bright and dark exciton states provides a measure of the spin-orbit splitting of the conduction band $\Delta_c = E_g^B - E_g^D$, where $E_g^B$ and $E_g^D$ denote, correspondingly, the single particle bandgaps associated with upper and lower conduction band subbands. It is worth noting, that the energy difference $\Delta_{DB}$ between the ground states of dark and bright excitons depends also on their respective, $E_b^D$ and $E_b^B$, binding energies: $\Delta_{DB} = \Delta_c + (E_b^D - E_b^B)$. Since $E_b^D$ and $E_b^B$ are *a priori* different, measuring the $\Delta_{DB}$ parameter alone does not infer the amplitude of $\Delta_c$.

Experimental results and discussion

The active part of the structure used for the experiments is a WSe$_2$ monolayer extracted from a commercially available tungsten disulfide crystal. The monolayer was encapsulated in between hBN flakes and deposited on a Si substrate, following the conventional methods of mechanical exfoliation and deterministic dry transfer techniques. Experiments consisted of low temperature (4.2 K) magneto-photoluminescence measurements carried out in magnetic fields up to 30 T, applied along the monolayer plane. As sketched in the inset to Fig. 2, we used the "back scattering" geometry where both the excitation and the collected light beams were quasi-perpendicular to the sample/monolayer. More details on sample preparation and experimental details can be found in SI.

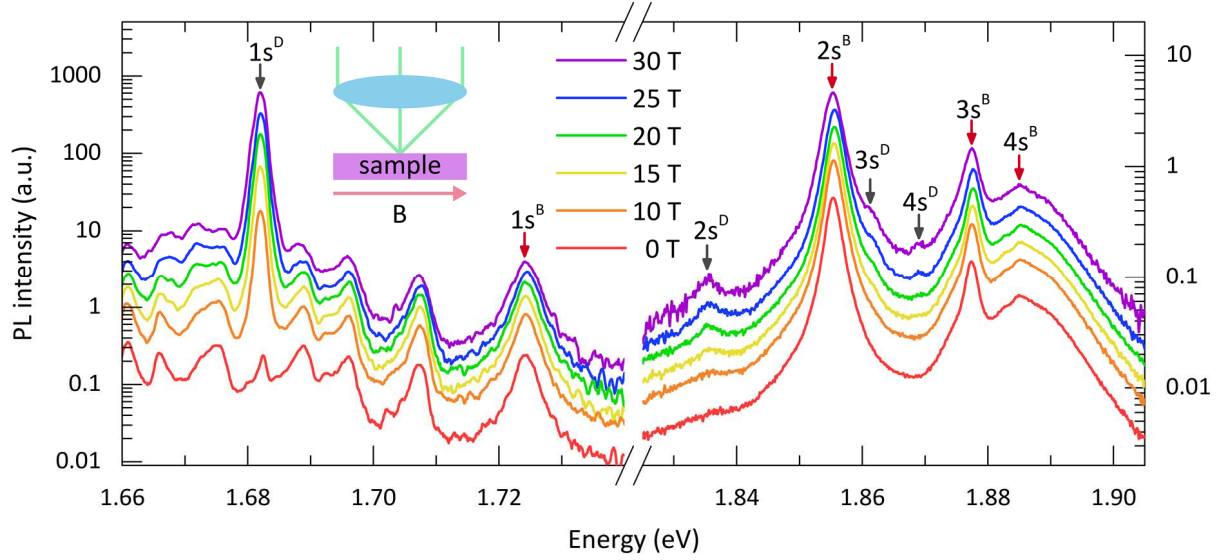

*Figure 2. Low-temperature photoluminescence spectra measured at selected in-plane magnetic fields in the energy region of both ground (left panel) and excited states (right panel). Spectra were shifted for clarity and normalized to the $1s^B$ feature in the lower energy region and to the $2s^B$ feature in the higher energy region. Note that the intensity scales differ for these two regions. The inset schematically shows the "back scattering" geometry used in the experiment, where the excitation and the collected light beams are quasi-perpendicular to the sample surface.*

Representative photoluminescence spectra of the investigated WSe$_2$ monolayer are presented in Fig.2. Inspecting the low spectral range (left panel of Fig.2), one recognizes a characteristic set of multiple photoluminescence transitions, typical of WSe$_2$ monolayers[31–33]. Possible formations of many different excitonic states (including nontrivial indirect/inter valley excitons and excitonic complexes such as trions and biexcitons) account for the spectrum complexity which we disregard here, focusing our attention on neutral, bright and dark intravalley excitons. First, we examine the photoluminescence (PL) spectrum measured in the absence of magnetic field (B), and, in accordance to previous reports, recognize the characteristic sequence of emission peaks related to the Rydberg series of $1s^B$, $2s^B$, $3s^B$, up to $4s^B$ states of bright excitons[24,25,27–29]. The PL spectrum measured at B=0 shows also a weak transition related to the ground state, $1s^D$, of dark exciton[14–16]. This is because of a legitimate emission from this state in the direction along the monolayer plane and therefore its weak visibility in our spectrum which we collected using the objective with relatively large numerical aperture[15,16]. Please also note, that we employ here a simplified picture of the dark exciton and view it as raising only a single resonance, i.e., we ignore its doublet, "dark-grey" components split by ~0.7 meV[14,16], an energy which is negligibly small in the context of our further considerations. Central for our report is the study of the evolution of the PL spectra measured as a function of the in-plane magnetic field ($B_\parallel$). Such studies have been already successfully applied to unveil the ground ($1s^D$) states of dark excitons in most of TMD monolayers[16–19,34]. The application of an in-plane magnetic field induces the Zeeman effect, and thus the mixing of spin-orbit split conduction subbands; dark excitons acquire an oscillator strength ($\sim B_\parallel^2$) for the emission in the direction perpendicular to the monolayer plane and become apparent in the PL spectra. The efficient "magnetic brightening" of the ground state of the dark exciton in WSe$_2$ monolayer is confirmed by the present experiments. As can be seen in the left panel of Fig.2, the intensity of the $1s^D$ emission grows progressively as a function of $B_\parallel$, and eventually dominates the PL spectrum in the limit of high magnetic fields. Importantly, new transitions driven by the application of the in-plane magnetic field appear also in the higher spectral range (see right panel of Fig.2) in the vicinity of the PL peaks associated to the excited states of bright excitons. These new transitions, labeled as $2s^D$, $3s^D$ and $4s^D$ in Fig.2, are attributed to the excited Rydberg states of the dark exciton.

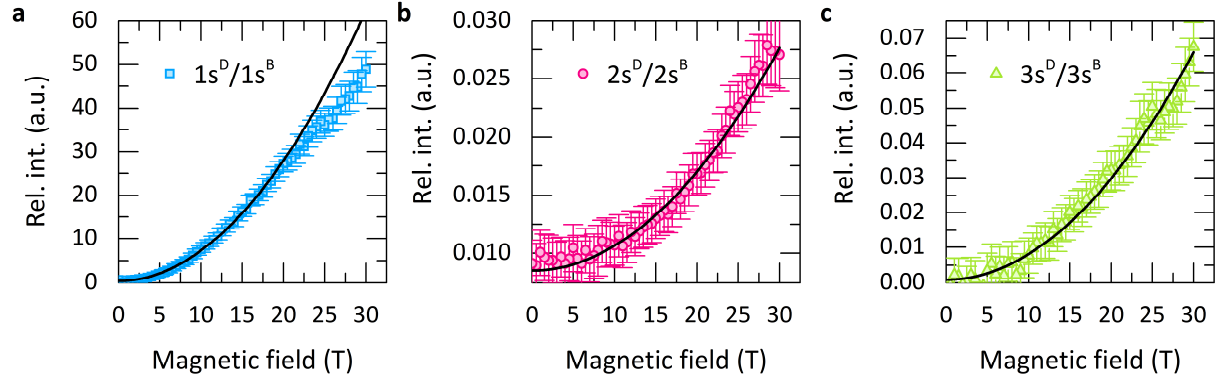

*Figure 3. Intensities of dark states relative to the intensity of bright states as a function of the in-plane magnetic field for **a** the 1s states (blue squares), **b** the 2s states (red circles) and **c** the 3s states (green triangles) along with quadratic fits (black lines) of the form $a + bx^2$. Note that in the case of 1s states only data points from 0 T to 20 T were considered for the fitting, as for higher fields, the relative intensity clearly starts to deviate from quadratic dependence.*

As demonstrated in Fig. 3, the magnetic brightening of dark exciton states, i.e. the increase of the relative intensity $I_D(n)/I_B(n)$ of dark $ns^D$ emission peak with respect to its bright $ns^B$ counterpart, roughly follows the expected $I_D(n)/I_B(n) = b_n B_\parallel^2$ rule (see SI for more details on data analysis). This, however, is not true in the case of ground states for the in-plane fields $B_\parallel > 20$ T, where the relative intensity deviates from the quadratic dependence, possibly due to the shortened (by optical activation) dark exciton lifetime leading to the decrease of the dark exciton population. Hence, only the datapoints for $B_\parallel < 20$ T were used for the estimation of $b_1$. Whereas the extracted values of $b_n$ are in reasonable agreement with the theoretical estimations for $n > 1$, the observed brightening of the ground dark $1s^D$ is by far more efficient ($b_1 \sim 10^4 b_n$, for $n > 1$) (see SI for more details). This might be due to the expected imbalance between populations of dark and bright exciton ground states, but the factor of $b_1/b_n \sim 10^4$ seems to be surprisingly large what points towards highly non-thermal population of optical excitations in a WSe$_2$ monolayer.

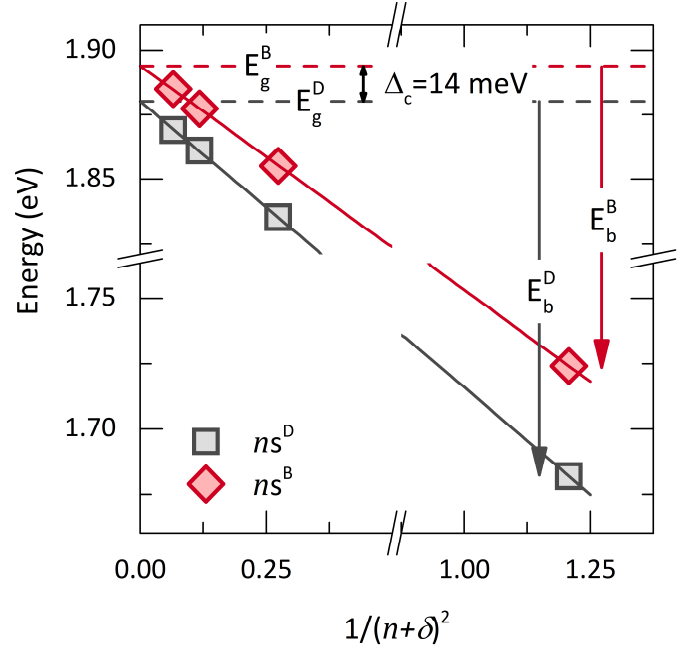

**Figure 4.** *Experimentally obtained transition energies for the bright (red, $ns^B$) and dark (grey, $ns^D$) exciton s-series as a function of $1/(n + \delta)^2$ where $\delta = -0.09$. The solid red and grey lines represent fits to the data with the model described in the text. The dashed red and grey lines denote the band gaps for dark ($E_g^D = 1.880$ eV) and bright ($E_g^B = 1.894$ eV) excitons, while the red and grey arrows show the binding energies of bright ($E_b^B = 170$ meV) and dark ($E_b^D = 198$ meV) excitons, respectively. The conduction band splitting ($\Delta_c = 14$ meV) is also indicated.*

Energy diagrams of both dark and bright exciton series are of our particular interest. Central positions of of $1s^D$ and $1s^B$ emission peaks can be directly read from the measured spectra with a reasonable precision, whereas the central energies of transitions associated with excited excitonic states have been extracted by fitting the high energy part of the spectra with multiple Lorentzian resonances (see SI section). As shown in Fig. 4, both series of bright and dark excitonic resonances follow the previously evoked (for bright states) semi-empirical formula[25]:

$$E_n^{D/B} = E_g^{D/B} - \frac{Ry^{D/B}}{(n+\delta)^2}, \qquad (1)$$

where $E_g^{D/B}$ is the "dark/bright" bandgap associated to the lower/upper conduction band subband and, $Ry^{D/B}$ should be, in the first approximation, identified with the effective Rydberg energy for the dark/bright exciton; $Ry^{D/B} = 13.6$ eV $\cdot \mu^{D/B}/(\varepsilon^2 m_0)$, where $\mu^{D/B}$ is the reduced effective mass embracing the valence band hole and electron from upper/lower conduction subbands, $\varepsilon$ is the dielectric constant of the surrounding hBN material and $m_0$ stands for the electron mass. $\delta$ parameter is found to be $\delta = -0.09$, which is notably common for dark and bright series and well matches the previously reported value ($-0.08$ for the bright exciton series[25]). The extracted values of "dark" and "bright" (single particle) bandgaps are, correspondingly, $E_g^D = 1.880$ eV and $E_g^B = 1.894$ eV and we eventually come up with disclosing the amplitude of the conduction band spin-orbit splitting: $\Delta_c = E_g^B - E_g^D = 14$ meV. Notably, this value is to be accounted by theoretical modeling of electronic bands of TMD monolayers[35–37] which, as yet, overestimates it by a factor of two. Binding energies of bright and dark excitons are two other extracted parameters: $E_b^B = 170$ meV and $E_b^D = 198$ meV, respectively. Following up our analysis (see Eq.1 and Fig.4), the resulting energy ratio $\rho_{DB} = E_b^D/E_b^B = 1.16$ (or difference $\Delta_{DB} = E_b^D - E_b^B = 28$ meV) is anticipated to be explained by different, for bright and dark excitons, reduced effective masses. ($E_b^D/E_b^B = \mu^D/\mu^B$). Indeed, when assuming the theoretically calculated masses of a valence hole ($m_h = 0.36$) and electrons in two conduction subbands ($m_e^D = 0.40$ and $m_e^B = 0.29$)[37], we find $\mu^D/\mu^B = 1.18$, a value which is very close to the experimentally derived ratio $E_b^D/E_b^B = 1.16$. On the other hand, one may expect that the difference between $E_b^D$ and $E_b^B$ arises not only from the direct Coulomb term but also from the exchange term, as bright and dark excitons consist of parallel or antiparallel spin configurations, respectively[38–40]. Whereas we suggest here that the exchange term might be relatively small, this claim has be to taken with caution, since our estimations refer to theoretically calculated effective masses which perhaps are not sufficiently accurate (as inaccurate as the amplitude of the spin orbit splitting). Nonetheless, our findings once again raise a question about the importance of the exchange interaction for the energy difference between the bright and dark excitons binding energies in TMD monolayers.

Conclusions

In this report we demonstrated the magnetic brightening of the Rydberg $ns$-series of dark excitons, up to $n = 4$, in a WSe$_2$ monolayer encapsulated in hBN. The analysis of the bright and dark excitons series allowed us to determine one of the missing band parameters, the amplitude of the spin-orbit splitting in the conduction band. Its derived value, $\Delta_c = 14$ meV, is significantly lower than commonly assumed, what calls for revision of theoretical calculations of electronic bands in TMD monolayers . Moreover, our results suggest that the difference between the binding energies of bright and dark excitons can be fully explained by the difference in the masses of electrons in the two spin-orbit-split conduction bands, without referring to exchange interactions.

Acknowledgements

The work has been supported by: EU Graphene Flagship, CNRS via IRP "2D materials", the ATOMOPTO project (TEAM programme of the Foundation for Polish Science, co-financed by the EU within the ERDFund), the Nano fab facility of the Institut Néel/CNRS/UGA, EMFL and DIR/WK/2018/07 grant from MEiN of Poland. M. B. acknowledges the financial support from the ERC under the European Union's Horizon 2020 research and innovation programme (GA no. 714850) and of the Ministry of Education, Youth and Sports of the Czech Republic under the project CEITEC 2020 (Grant No. LQ1601). K.W. and T.T. acknowledge support from the Elemental Strategy Initiative conducted by the MEXT, Japan, (grant no. JPMXP0112101001), JSPS KAKENHI (grant no. JP20H00354), and the CREST (JPMJCR15F3), JST.

References

1. Bayer, M. *et al.* Fine structure of neutral and charged excitons in self-assembled In(Ga)As/(Al)GaAs quantum dots. *Phys. Rev. B* **65**, 195315 (2002).
2. Slobodeniuk, A. O. & Basko, D. M. Spin-flip processes and radiative decay of dark intravalley excitons in transition metal dichalcogenide monolayers. *2D Mater.* **3**, 35009 (2016).
3. Combescot, M., Combescot, R. & Dubin, F. Bose–Einstein condensation and indirect excitons: a review. *Reports Prog. Phys.* **80**, 66501 (2017).
4. Glasberg, S., Shtrikman, H., Bar-Joseph, I. & Klipstein, P. C. Exciton exchange splitting in wide GaAs quantum wells. *Phys. Rev. B* **60**, R16295–R16298 (1999).
5. Bayer, M., Stern, O., Kuther, A. & Forchel, A. Spectroscopic study of dark excitons in InxGa1–xAs self-assembled quantum dots by a magnetic-field-induced symmetry breaking.

*Phys. Rev. B* **61**, 7273–7276 (2000).
6. Kowalik, K. *et al.* Manipulating the exciton fine structure of single CdTe/ZnTe quantum dots by an in-plane magnetic field. *Phys. Rev. B* **75**, 195340 (2007).
7. Zaric, S. *et al.* Optical Signatures of the Aharonov-Bohm Phase in Single-Walled Carbon Nanotubes. *Science* **304**, 1129–1131 (2004).
8. Shaver, J. *et al.* Magnetic Brightening of Carbon Nanotube Photoluminescence through Symmetry Breaking. *Nano Lett.* **7**, 1851–1855 (2007).
9. Srivastava, A., Htoon, H., Klimov, V. I. & Kono, J. Direct Observation of Dark Excitons in Individual Carbon Nanotubes: Inhomogeneity in the Exchange Splitting. *Phys. Rev. Lett.* **101**, 87402 (2008).
10. Goryca, M. *et al.* Brightening of dark excitons in a single CdTe quantum dot containing a single $Mn^{2+}$ ion. *Phys. Rev. B* **82**, 165323 (2010).
11. Alexander-Webber, J. A. *et al.* Hyperspectral Imaging of Exciton Photoluminescence in Individual Carbon Nanotubes Controlled by High Magnetic Fields. *Nano Lett.* **14**, 5194–5200 (2014).
12. Zhou, Y. *et al.* Probing dark excitons in atomically thin semiconductors via near-field coupling to surface plasmon polaritons. *Nat. Nanotechnol.* **12**, 856–860 (2017).
13. Zhang, X.-X. *et al.* Magnetic brightening and control of dark excitons in monolayer $WSe_2$. *Nat. Nanotechnol.* **12**, 883–888 (2017).
14. Robert, C. *et al.* Fine structure and lifetime of dark excitons in transition metal dichalcogenide monolayers. *Phys. Rev. B* **96**, 1–8 (2017).
15. Wang, G. *et al.* In-Plane Propagation of Light in Transition Metal Dichalcogenide Monolayers: Optical Selection Rules. *Phys. Rev. Lett.* **119**, 1–7 (2017).
16. Molas, M. R. *et al.* Probing and Manipulating Valley Coherence of Dark Excitons in Monolayer $WSe_2$. *Phys. Rev. Lett.* **123**, 96803 (2019).
17. Lu, Z. *et al.* Magnetic field mixing and splitting of bright and dark excitons in monolayer $MoSe_2$. *2D Mater.* **7**, 15017 (2019).
18. Robert, C. *et al.* Measurement of the spin-forbidden dark excitons in $MoS_2$ and $MoSe_2$ monolayers. *Nat. Commun.* **11**, 4037 (2020).
19. Zinkiewicz, M. *et al.* Neutral and charged dark excitons in monolayer $WS_2$. *Nanoscale* **12**, 18153–18159 (2020).
20. Farenbruch, A., Fröhlich, D., Yakovlev, D. R. & Bayer, M. Rydberg Series of Dark Excitons in $Cu_2O$. *Phys. Rev. Lett.* **125**, 207402 (2020).
21. Watanabe, K., Uchida, K. & Miura, N. Magneto-optical effects observed for GaSe in megagauss magnetic fields. *Phys. Rev. B* **68**, 155312 (2003).
22. Nagamune, Y., Takeyama, S. & Miura, N. Exciton spectra and anisotropic Zeeman effect in $PbI_2$ at high magnetic fields up to 40 T. *Phys. Rev. B* **43**, 12401–12405 (1991).
23. Chernikov, A. *et al.* Exciton binding energy and nonhydrogenic Rydberg series in monolayer $WS_2$. *Phys. Rev. Lett.* **113**, (2014).
24. Stier, A. V. *et al.* Magnetooptics of Exciton Rydberg States in a Monolayer Semiconductor. *Phys. Rev. Lett.* **120**, 57405 (2018).
25. Molas, M. R. *et al.* Energy Spectrum of Two-Dimensional Excitons in a Nonuniform Dielectric Medium. *Phys. Rev. Lett.* **123**, 136801 (2019).
26. Goryca, M. *et al.* Revealing exciton masses and dielectric properties of monolayer semiconductors with high magnetic fields. *Nat. Commun.* **10**, 4172 (2019).
27. Liu, E. *et al.* Magnetophotoluminescence of exciton Rydberg states in monolayer $WSe_2$. *Phys. Rev. B* **99**, 205420 (2019).
28. Chen, S.-Y. *et al.* Luminescent Emission of Excited Rydberg Excitons from Monolayer $WSe_2$. *Nano Lett.* **19**, 2464–2471 (2019).
29. Delhomme, A. *et al.* Magneto-spectroscopy of exciton Rydberg states in a CVD grown $WSe_2$ monolayer. *Appl. Phys. Lett.* **114**, 232104 (2019).
30. Delhomme, A. *et al.* Flipping exciton angular momentum with chiral phonons in $MoSe_2/WSe_2$

heterobilayers. *2D Mater.* **7**, 41002 (2020).
31. Li, Z. *et al.* Emerging photoluminescence from the dark-exciton phonon replica in monolayer WSe2. *Nat. Commun.* **10**, 2469 (2019).
32. Liu, E. *et al.* Valley-selective chiral phonon replicas of dark excitons and trions in monolayer WSe2. *Phys. Rev. Res.* **1**, 32007 (2019).
33. He, M. *et al.* Valley phonons and exciton complexes in a monolayer semiconductor. *Nat. Commun.* **11**, 618 (2020).
34. Molas, M. R. *et al.* Brightening of dark excitons in monolayers of semiconducting transition metal dichalcogenides. *2D Mater.* **4**, 21003 (2017).
35. Komider, K., González, J. W. & Fernández-Rossier, J. Large spin splitting in the conduction band of transition metal dichalcogenide monolayers. *Phys. Rev. B - Condens. Matter Mater. Phys.* **88**, 1–7 (2013).
36. Liu, G. Bin, Shan, W. Y., Yao, Y., Yao, W. & Xiao, D. Three-band tight-binding model for monolayers of group-VIB transition metal dichalcogenides. *Phys. Rev. B - Condens. Matter Mater. Phys.* **88**, (2013).
37. Kormányos, A. *et al.* kp theory for two-dimensional transition metal dichalcogenide semiconductors. *2D Mater.* **2**, 22001 (2015).
38. Echeverry, J. P., Urbaszek, B., Amand, T., Marie, X. & Gerber, I. C. Splitting between bright and dark excitons in transition metal dichalcogenide monolayers. *Phys. Rev. B* **93**, 121107 (2016).
39. Deilmann, T. & Thygesen, K. S. Dark excitations in monolayer transition metal dichalcogenides. *Phys. Rev. B* **96**, 201113 (2017).
40. Bieniek, M., Szulakowska, L. & Hawrylak, P. Band nesting and exciton spectrum in monolayer MoS2. *Phys. Rev. B* **101**, 125423 (2020).

# Supplementary Information:
# Rydberg series of dark excitons and the conduction band spin-orbit splitting in monolayer $WSe_2$

## 1. Methods

*Sample fabrication.* The sample used in our study is composed of $WSe_2$ monolayer encapsulated in hBN flakes and deposited on a Si substrate. The heterostructure was fabricated using mechanical exfoliation of bulk $WSe_2$ and hBN crystals. The bottom hBN layer was created by non-deterministic exfoliation onto Si substrate, while the $WSe_2$ monolayer and capping hBN flake were transferred using dry stamping technique onto bottom hBN flake using PDMS stamps.

*Experimental setup.* The low-temperature magneto-photoluminescence experiments were performed in the Voigt configuration, with magnetic field applied parallel to the monolayer plane using free-beam insert placed in a resistive solenoid producing magnetic fields up to 30 T. The sample was placed on top of the x-y-z piezo stage and kept in gaseous helium at T=4.2 K. CW 515 nm excitation was used, with the excitation power of 200 µW. The excitation beam was focused and the signal was collected by the same microscope objective with a numerical aperture of NA=0.35. The collected light was analyzed by a 0.5 m long monochromator equipped with a 500 lines/mm grating and CCD camera.

## 2. Emission intensities and fitting procedure

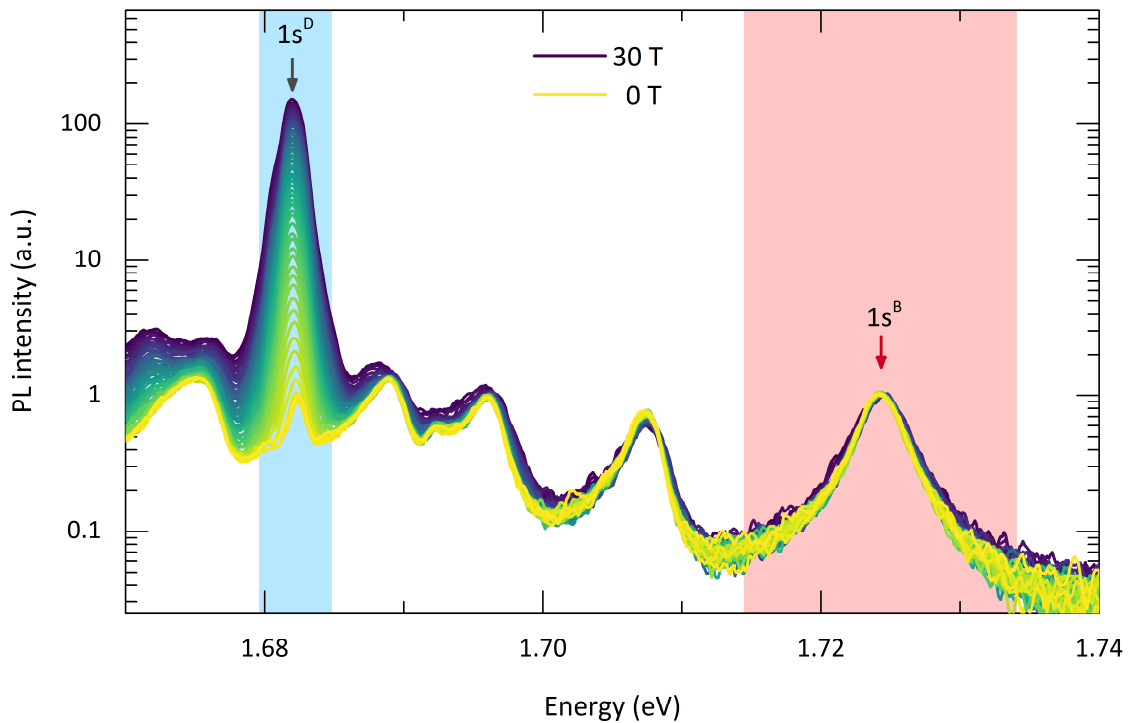

**Figure S1.** *The low-temperature photoluminescence spectra measured at in-plane magnetic fields from 30 T to 0 T in the energy region of ground states. The spectra were normalized to $1s^B$ feature for clarity. The red/blue rectangles highlight the regions that were integrated to obtain the total PL intensity of $1s^B/1s^D$.*

Quantitative analysis of the brightening effect of dark states by the in-plane magnetic fields requires knowledge of the intensity of emission of both bright and dark states. To obtain such intensities of ground exciton $1s^B$ and $1s^D$ states we used area/integration method. In the Fig. S1 we show the regions used for integrations. As the ground bright state $1s^B$ is well separated from the other features, it is relatively easy to define integration region. In the case of ground dark state $1s^D$ however, the emission line partially overlaps with neighbouring transitions, inflicting non-negligible error in our estimation. Hence, two regions were integrated to obtain the error of $1s^D$ intensity – one that fully covers the emission line at 30 T, and the other that covers only $1s^D$ at 0 T. The error $1s^D$ intensity estimation is then calculated as the difference between these two areas.

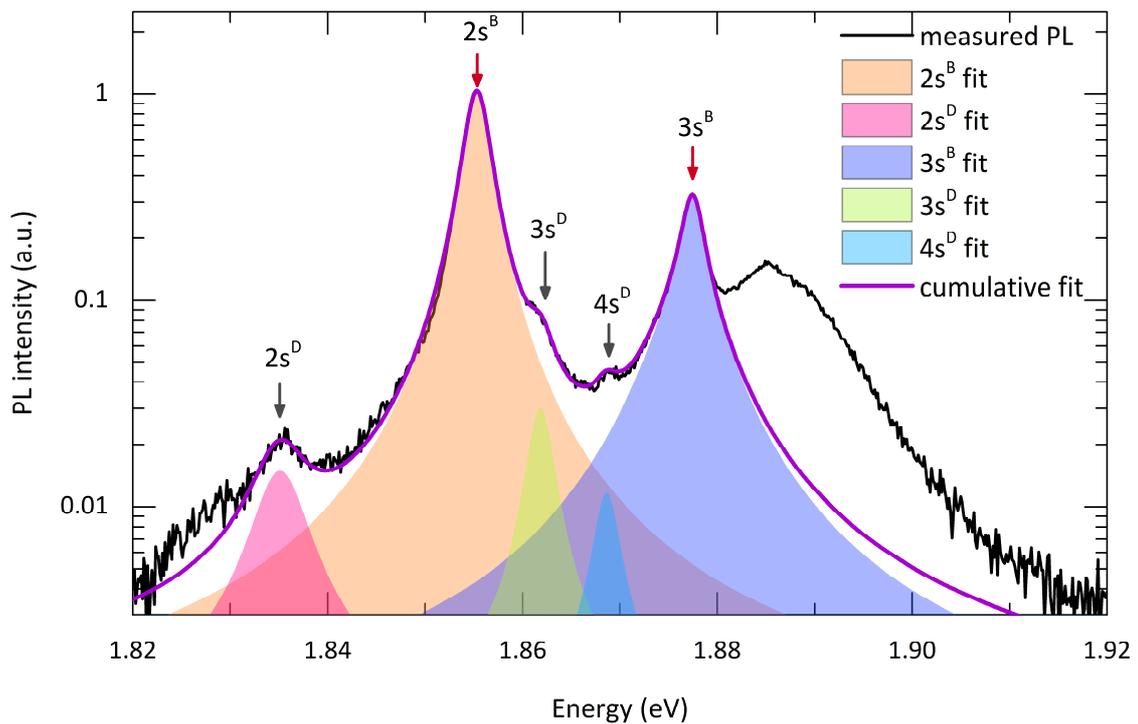

*Figure S2.* *The low-temperature photoluminescence spectrum measured at 30 T in the energy region of excited states, along with the fitted curves. The measured spectrum is represented by black line, the intensities of excited bright and dark states are represented by colored areas, while cumulative fit is represented by purple line. The same energies and linewidths of features were used to fit the spectra obtained at lower magnetic fields.*

Similar approach could not be applied in the case of the excited bright and dark exciton states, as their emission lines significantly overlap. Hence the spectrum at 30 T was fitted in the region of interest using Lorentzian functions for the emission lines, allowing us to get nice agreement between measured and fitted spectrum, see Fig. S2, and extract the areas of the emission lines. The same energies and linewidths of the Lorentzian functions were then used to fit the spectra at lower magnetic fields.

3. Brightening strength

The obtained intensities of bright and dark excitonic states were used to calculate the relative intensity of dark states to bright states as a function of magnetic field, which can be considered a measure of the brightening effect. In the Fig. 3 of the main text we show relative intensity evolution for 1s, 2s and

3s states, along with the fitted parabolas ($a + bx^2$). As for the 1s states this evolution starts to deviate from quadratic above 20 T, only the data for 0-20 T are taken into account during fitting. We obtained following parameters: $b_1^E = (69 \pm 6)10^{-2}$, $b_2^E = (21 \pm 6)10^{-6}$ and $b_3^E = (7 \pm 2)10^{-5}$.

These parameters can be estimated theoretically, when the occupation difference of the states is ignored, using formula[1]:

$$b_n^T = \left[\frac{g_\parallel \mu_B}{2(E_n^B - E_n^D)}\right]^2,$$

where $g_\parallel$ is the in-plane g-factor, $\mu_B$ is the Bohr magneton, $B_\parallel$ represents the in-plane magnetic field, while $E_n^{B/D}$ denote the zero-field energies of bright and dark states, respectively, for given principal quantum number $n$. Using the $g_\parallel = 2.7$[1] and the energies obtained from our experiments, we obtained parameters: $b_1^T = 34 \cdot 10^{-7}$, $b_2^T = 15 \cdot 10^{-6}$ and $b_3^T = 23 \cdot 10^{-6}$. Whereas in the case the excited states ($n = 2,3$) there is reasonable agreement between values obtained theoretically and experimentally, for the ground dark state the parameters differ by 5 orders of magnitude. Such enormous difference cannot be explained simply by the thermal population difference, suggesting highly non-thermal population of optical excitations in a WSe$_2$ monolayer.

References

1. Molas, M. R. *et al.* Probing and Manipulating Valley Coherence of Dark Excitons in Monolayer WSe2. *Phys. Rev. Lett.* **123**, 96803 (2019).